# Comparative analysis of large data processing in Apache Spark using Java, Python and Scala[⋆]

Ivan Borodii[1,∗,†], Illia Fedorovych[1,∗,†], Halyna Osukhivska[1,∗,†], Diana Velychko[2,∗,†] and Roman Butsii[3,∗,†]

[1] *Ternopil Ivan Puluj National Technical University, 56 Ruska St, Ternopil, UA46001, Ukraine*
[2] *Rochester Institute of Technology, 1 Lomb Memorial Dr, Rochester, New York 14623, USA*
[3] *Institute of Telecommunications and Global Information Space of the National Academy of Sciences of Ukraine, 13 Chokolivskiy blvd., Kyiv, UA02000, Ukraine*

**Abstract**
During the study, the results of a comparative analysis of the process of handling large datasets using the Apache Spark platform in Java, Python, and Scala programming languages were obtained. Although prior works have focused on individual stages of Spark processing or specific storage formats, comprehensive comparisons of full ETL workflows across programming languages using Apache Iceberg remain limited. This study addresses this gap. The analysis was performed by executing several operations, including downloading data from CSV files, transforming it, and loading it into an Apache Iceberg analytical table. It was found that the performance of the Spark algorithm varies significantly depending on the amount of data and the programming language used. In particular, when processing a 5-megabyte CSV file, the best result was achieved in Python – 6.71 seconds, which is superior to Scala's score of 9.13 seconds and Java's time of 9.62 seconds. At the same time, for processing a large CSV file of 1.6 gigabytes, all programming languages demonstrated similar results: the fastest performance was demonstrated in Python – 46.34 seconds, while Scala and Java showed results of 47.72 and 50.56 seconds, respectively. When performing a more complex operation that involved combining two CSV files into a single dataset for further loading into an Apache Iceberg table, Scala demonstrated the highest performance, at 374.42 seconds. Java processing was completed in 379.8 seconds, while Python was the least efficient, with a runtime of 398.32 seconds. It follows that the programming language significantly affects the efficiency of data processing by the Apache Spark algorithm, with Scala and Java being more productive for processing large amounts of data and complex operations, while Python demonstrates an advantage in working with small amounts of data. The results obtained can be useful for optimizing data handling processes depending on specific performance requirements and the amount of information being processed. The observed trends provide practical guidelines for selecting the optimal programming language, considering data volume, operational complexity, and runtime characteristics, which can inform the design of efficient large-scale data processing pipelines.

## 1. Introduction

Nowadays, processing large amounts of data is one of the key challenges as the volume of data is constantly growing. This applies to all types of data generated in social media, the Internet of Things (IoT), intelligent systems, or business processes. In such circumstances, it is critical to develop and improve tools that ensure efficient and reliable processing, analysis, and management of big data. The most well-known data processing tool is Apache Spark, which provides high-performance parallel execution on clusters, accelerating work with large data sets in a distributed environment. Apache Spark supports in-memory processing, which significantly reduces the execution time compared to traditional disc-based computing, such as Hadoop MapReduce [1]. A key feature of Apache Spark is its support for multiple programming languages, including Java, Python, Scala, and R. Each language has its unique characteristics that impact the speed of algorithm execution, resource efficiency, and ease of development. For example, Scala is the native language for Apache Spark because the platform itself is written in Scala, which provides tight integration and access to low-level functions. Java is one of the oldest languages supported by Spark, noted for its fast execution speed. In contrast, Python is popular due to its simple syntax and extensive library ecosystem.

The choice of Apache Spark environment can have a significant impact on system performance, particularly when processing large datasets. Even slight differences in performance can have a significant impact in complex systems, where a few seconds of delay can translate into hours of additional processing time. Additionally, the choice of language can impact aspects such as memory consumption, task stability, and scalability.

Research on the impact of programming languages on the performance of the full data processing cycle in modern analytical systems has not received much attention in the scientific literature. Therefore, a comprehensive comparison of the performance of ETL processes in Apache Spark and Apache Iceberg, using Java, Python, and Scala programming languages, with the same Spark configurations and input datasets, is an urgent task for processing large datasets.

## 2. Related works

Several researchers have studied the performance of Apache Spark for big data processing. In particular, Vergas' study [2] compared Apache Iceberg with Delta Lake and Apache Hudi on a cloud object store. The testing was conducted in the Apache Spark environment using the Scala language, which allowed the full use of the Catalyst optimizer extension for each format. Iceberg demonstrated the highest performance in the data reading phase, with a result up to 20% faster than Hudi and 10-15% faster than Delta Lake. However, in the update phase, Iceberg was twice as slow as Delta Lake. The study concluded that Iceberg is the most suitable for implementing analytical data structures.

In Ciesielski's work [3], Apache Iceberg was used in a Domino financial application based on Apache Spark. The data was processed using PySpark, resulting in the efficient processing of large tables with high consistency. By using ACID transactions and versioning, Iceberg provided a reduction in data access time. In this work, Iceberg was effective for tasks where version control and data integrity are essential.

Research [4] compared Apache Iceberg with ORC, Parquet, and Avro in Apache Spark, Hive, and Hadoop environments, using both Java and Scala. Testing was conducted on a cloud infrastructure, with a focus on query speed. Iceberg provided up to 30% better performance when running analytical queries compared to ORC and Parquet, while also offering high stability and flexibility in scaling. The authors recommend it for cloud systems focused on large analytical workloads.

The study [5] compared the continuous and microbatching modes of Spark Structured Streaming using various configurations and tests, focusing on latency and throughput. It also describes the practical implications of deploying continuous processing mode in real-world applications, evaluating its compatibility with various data sources and receivers, mainly Apache Kafka. In particular, the continuous processing mode offers significantly lower latency when using the Rate source (2 ms instead of 528 ms in microbatch mode); its performance in high-throughput scenarios using the Kafka source may be less consistent (260 ms as opposed to 197 ms in microbatch mode). This is crucial for optimizing text data flow strategies in big data analytics, providing information that can inform the selection of processing modes tailored to specific operational needs. Unlike previous studies that have mostly focused on individual languages or stages of data processing, this study implements a unified approach to testing all process components required by the ETL concept: data selection, data transformation, and storage in Apache Iceberg format.

In the study [6], Spark was compared with Hadoop MapReduce on a cluster of 10 nodes, 80 cores, and 60 TB of data storage. Testing was conducted using the WordCount and TeraSort workloads to evaluate performance when processing data up to 600 GB. Processing parameters, such as partitioning volume, buffer values, and shuffle file size, were adjusted to investigate their impact on performance. Compared to Hadoop, Apache Spark was twice as fast when processing WordCount and up to 14 times faster when processing TeraSort. All implementations were written in Scala, which allowed the use of native Spark APIs and ensured maximum compatibility with the system kernel. The authors emphasize that with optimized configurations, Spark outperforms Hadoop in most analytical processing scenarios.

A scientific study [7] compared Hadoop MapReduce and Apache Spark based on two types of computing loads: iterative and streaming. Spark demonstrated 2-10 times higher performance in tasks with data repetition due to in-memory processing. The testing was conducted using all supported Spark languages, including Java, Python, and Scala, which enabled the evaluation of API usability and performance in each case. Specifically, Scala was used for high-performance scenarios, Python for integration with machine learning libraries, and Java as a basic option. Spark also proved to be more efficient in implementing SQL queries and processing data streams without the need for external tools.

In the research [8], Spark was compared with Hadoop MapReduce in the context of architecture, APIs, graph processing capabilities, machine learning, and SQL analytics. Testing was conducted using Spark components: Spark SQL, Spark Core, SparkR, GraphX, MLlib, and data received from Twitter, Kafka, and IoT sensors. It was demonstrated that Spark processes data significantly faster due to its in-memory computing capabilities and improved integration with modern tools. Python, Scala, and R programming languages were used to implement the computations. SparkR was utilized for analyzing large amounts of data within the R environment, while Python and Scala were employed for the primary components of analytics and real-time data processing. The authors note Spark as a cross-platform solution with high real-time performance.

## 3. Proposed methodology

This study compares the performance of Apache Spark for processing large datasets in three popular programming languages: Java, Python, and Scala. The input data for processing were obtained from the open-meteo web resource [9], which contains hourly atmospheric air temperature data for 2024, depending on geographical coordinates. These data were downloaded to a 1.6 GB CSV file. The study also utilized another input CSV file containing a list of cities and their corresponding geographical coordinates, which totaled approximately 5 megabytes and was obtained from the web resource simplemaps.com [10]. The coordinate information was used to merge the two CSV files, forming a dataset with hourly air temperature data for each city in 2024. All the data was loaded into Apache Iceberg tables, which enables efficient management of large amounts of structured data. This process falls under the concept of ETL (Extract, Transform, Load), which involves migrating data from one structure to another [11]. The ETL process includes three main stages: extraction, transformation, and loading. In the context of this study, the extraction stage involves reading data from input CSV files. At the transformation stage, the data is modified following the research requirements using the Spark algorithm. The loading procedure involves migrating the prepared data to Apache Iceberg tables, a modern format for storing large amounts of data designed to ensure stability, scalability, and ease of integration with various analytical platforms [12]. Iceberg enables the creation of tables with a multi-level organization of sections, ensuring query optimization and increasing the productivity of operations with large amounts of data. Another essential feature of Iceberg is its support for data evolution schemas. This ensures that the structure of tables can be modified without the need to rewrite the entire database [13].

The analysis consisted of three parts. The first stage involved loading the list of cities into an Iceberg table. The second stage involved importing a file containing hourly temperature readings by coordinates. The third stage was the most complex, involving the merging of two files by coordinates and loading the resulting dataset into a separate Iceberg table. In this case, the table format was structured into data partitions by date and country, which significantly affected the execution time. Loading data into Iceberg at this stage took significantly more time compared to the previous two stages. The test data was prepared using two main data sources. The first source was a CSV file containing a list of cities, each with its own set of geographic coordinates. The file was downloaded from the web resource simplemaps.com, which contains a list of about 47 thousand cities. Table 1 shows an example of how the data in this file is displayed.

**Table 1**
A list of cities retrieved from simplemaps.com

| City    | Cityascii | Lat   | Lng    | Country   | Iso2 | Admin   | Capital | Population |
|---------|-----------|-------|--------|-----------|------|---------|---------|------------|
| Tokyo   | Tokyo     | 35.7  | 139.7  | Japan     | JP   | Tōkyō   | primary | 37732000   |
| Jakarta | Jakarta   | -6.17 | 106.8  | Indonesia | ID   | Jakarta | primary | 33756000   |
| Delhi   | Delhi     | 28.6  | 77.23  | India     | IN   | Delhi   | admin   | 135687260  |

The second source was the open-meteo web resource, which was used to collect weather data for 2024 in the form of hourly records. It includes information on temperature, date, time, and geographical coordinates. The total file size is about 1.6 GB. Table 2 provides an example of the data contained in the file.

**Table 2**
A list of hourly temperature data retrieved from open-meteo.com

| Date                      | Temperature_2m | Lat     | Lng      |
|---------------------------|----------------|---------|----------|
| 2024-01-01 00:00:00+00:00 | 8.6            | 35.6897 | 139.6922 |
| 2024-01-01 01:00:00+00:00 | 9.6            | 35.6897 | 139.6922 |
| 2024-01-01 02:00:00+00:00 | 9.7            | 35.6897 | 139.6922 |

For the purposes of testing on Python, Scala, and Java programming languages, three separate database schemas were created for each language: worldcities_java, worldcities_scala, and worldcities_python. Each of these schemas contains three tables that correspond to different aspects of data analysis. The weather_events table stores information about weather events obtained from a CSV file generated after downloading data from the open-meteo web resource. The table is divided into sections by date. The weather_events entity contains the attributes shown in Table 3.

The world_cities table contains information about countries, cities, provinces, population, coordinates, and other relevant details. It receives data from the simplemaps.com website in the format of a CSV file. The table contains approximately 47,000 records and is divided into sections by country in ISO 2 format [14]. The world_cities entity contains the attributes shown in Table 4.

**Table 3**
Weather_events table description

| Field name     | Data type | Partitioning    |
|----------------|-----------|-----------------|
| lat            | double    | Not partitioned |
| lng            | double    | Not partitioned |
| temperature_ts | timestamp | Not partitioned |
| temperature    | double    | Not partitioned |
| temperature_dt | date      | Partitioned     |

The weather_hourly_events entity combines data from two test CSV files. The merging is done using lat (latitude), lng (longitude) fields that exist in both files. The table is divided into partitions by date and country. The list of attributes of the weather_hourly_events entity is shown in Table 5.

**Table 4**
World_cities table description

| Field name | Data type | Partitioning |
| --- | --- | --- |
| country | string | Not partitioned |
| province | string | Not partitioned |
| city | string | Not partitioned |
| capital | string | Not partitioned |
| population | bigint | Not partitioned |
| lat | double | Not partitioned |
| lng | double | Not partitioned |
| country_iso2 | string | Partitioned |

**Table 5**
Weather_hourly_events table description

| Field name | Data type | Partitioning |
| --- | --- | --- |
| lat | double | Not partitioned |
| lng | double | Not partitioned |
| temperature_ts | timestamp | Not partitioned |
| temperature | double | Not partitioned |
| population | bigint | Not partitioned |
| country | string | Not partitioned |
| province | string | Not partitioned |
| city | string | Not partitioned |
| temperature_dt | date | Partitioned |
| country_iso2 | string | Partitioned |

The computational experiment involving the launch of Spark applications was conducted on a computer running the Windows 10 Home (64-bit) operating system with 16 GB of RAM and an AMD Ryzen 5 4600H processor, accompanied by a Radeon Graphics card (3.00 GHz). This hardware and software configuration provided an environment for evaluating system performance.

The experiment used Apache Spark 3.5.1 for distributed data processing and Apache Iceberg 1.7.1 for managing analytical tables. The software environment included Java 11.0.14, Scala 2.12.18, and Python 3.11. Integrated development environments, such as PyCharm Professional Edition for Python and IntelliJ IDEA Community Edition for Scala and Java, were used for code development and execution.

On the computer used for the experiment, the JVM heap size was approximately 3934 MB, with up to 12 parallel tasks running on 12 cores. All tasks share this memory, so excessive use could result in garbage collection or out-of-memory errors, which would reduce performance. Fine-tuning parallelism and memory usage is critical for stable execution.

A Spark session is the main entry point for interacting with Apache Spark. It is responsible for managing resources, executing queries, and interacting with files and databases. Initialization is a mandatory operation in every Spark application, as it determines the session's configuration, connections to data stores, and resource allocation [15].

The Spark configurations are defined in the same way for each of the studied programming languages and are presented in Table 6.

As part of the study, three separate experiments were conducted to implement the ETL process of loading data from CSV files into Apache Iceberg tables using Apache Spark in Java, Scala, and

Python. In each experiment, data is loaded from CSV files into a DataFrame (Python) or a Dataset (Java, Scala) using Spark. The code snippets for reading CSV files for each programming language are shown in Table 7.

**Table 6**
Spark configurations

| Configuration parameter | Description |
|---|---|
| master = local[*] | Enables running Spark application in local mode using a multithreading approach. |
| spark.sql.catalog.local | Defines the catalog for working with Iceberg tables. |
| spark.sql.catalog.local.type | Specifies the catalog type (Hadoop means file system-based storage). |
| spark.sql.catalog.local.warehouse | Sets the warehouse path for Iceberg data storage. |
| spark.sql.extensions | Enables Iceberg extensions in Spark. |
| spark.executor.instances = 5 | Uses 6 executors. |
| spark.executor.cores = 2 | Assigns 4 CPU cores per executor. |
| spark.driver.cores = 2 | Assigns 2 CPU cores to the driver. |
| spark.sql.adaptive.enabled = true | Enables adaptive query execution. |
| spark.sql.adaptive.coalescePartitions.enabled = true | Automatically reduces the number of small partitions. |
| spark.sql.adaptive.skewJoin.enabled = true | Optimizes skewed joins when data distribution is uneven. |
| spark.sql.optimizer.dynamicPartitionPruning = enabled | Reads only the required partitions during queries. |
| spark.memory.offHeap.enabled = true | Enables off-heap memory usage to reduce JVM garbage collection. |

**Table 7**
Reading CSV files by Spark for each programming language

| Language | Code line |
|---|---|
| Java | spark.read().option("header", "true").option("inferSchema", "true").csv(csvPath); |
| Scala | spark.read.option("header", "true").option("inferSchema", "true").csv(csvPath); |
| Python | spark.read.csv(csv_path, header=True, inferSchema=True) |

Setting the header=true and inferSchema=true parameters enables the data structure to be automatically detected and headers to be read correctly. After reading the CSV, the fields are selected and renamed for ease of further processing using the select() and alias() methods. At the transformation stage, operations are performed to bring the data to a consistent state. Such operations included the dropDuplicates() function for duplicate removal, the to_date() function for date conversion, and the join() function for dataset joining.

To optimize the join, the broadcast() function is used, which enables placing datasets in memory and accelerating query execution. However, it is effective only when the dataset contains a small volume of several megabytes that needs to be merged with a large dataset [16]. Table joins are performed using attributes that describe lat and lng coordinates, which allows you to integrate weather and geographic data into a single set.

Data is written to Iceberg tables using the overwrite mode, which provides for updating data without losing historical information. Using the overwrite-mode=dynamic parameter allows Spark to optimize the data update process and minimize the number of overwritten batches [17]. Fragments of the code for writing data to an Iceberg table are shown in Table 8.

**Table 8**
Writing data into Apache Iceberg tables by Spark for each programming language

| Language | Code line |
|---|---|
| Java | csvDF.write().format("iceberg").mode("overwrite").option("overwrite-mode", "dynamic").save(); |
| Scala | df.write.format("iceberg").mode("overwrite").option("overwrite-mode", "dynamic").save(); |
| Python | df.write.format("iceberg").mode("overwrite").option("overwrite-mode", "dynamic").save(); |

Storing data in the Apache Iceberg format ensures ACID transactions and efficient management of historical versions. This minimizes the risk of data loss and improves the performance of ETL processes [18].

## 4. Results of the experiment

The results of the Spark algorithm, which includes collecting data from CSV files, processing them, and loading them into Apache Iceberg tables for each of the studied programming languages, are shown in Table 9.

**Table 9**
Performance comparison between programming languages for running Spark algorithm in seconds

| Apache Iceberg table name | Java (sec) | Scala (sec) | Python (sec) |
|---|---|---|---|
| world_cities | 9.62 | 9.13 | 6.71 |
| weather_events | 50.56 | 47.72 | 46.34 |
| weather_hourly_events | 379.8 | 374.42 | 398.32 |

Data in Apache Iceberg is structured as a file system, consisting of partitions. This enables the efficient storage, transformation, and organization of large amounts of data by dividing it into separate sections based on specific characteristics, thereby enhancing performance during query execution [19].

Iceberg tables store data in Parquet files. This is a columnar data storage format widely used in data processing systems, such as Apache Spark, Hive, and Apache Flink. As Parquet is a columnar format, it stores data in columns rather than rows, which significantly increases the speed of processing large amounts of data, especially when working with individual table fields. This approach reduces the amount of data transferred from disk to memory, thereby improving query execution time. One of the primary advantages of Parquet files is their efficient data compression, which reduces the disk space occupied [20]. For each programming language under study, separate database schemas were created in the form of file directories. Each schema contains tables that are also organized as directories. The organization of Apache Iceberg schemas and tables is shown in Figures 1 and 2.

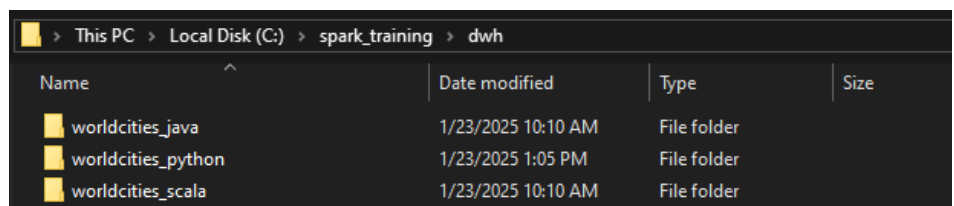

**Figure 1:** Schema organization of Apache Iceberg databases.

**Figure 2:** Table organization of Apache Iceberg schema.

Two main directories are created in each Apache Iceberg table: The data directory contains files with data in the Parquet format. Data is stored in further subdirectories, organized by table partitions. This enables efficient data storage and provides quick access to data during query execution [21]. The metadata directory contains information about the table schema, its batches, as well as the history of changes and transactions. Metadata is necessary for managing data in Iceberg, supporting ACID transactions, and other purposes [22].

The weather_hourly_events table contains partitions on the temperature_dt and country_iso2 fields, but since temperature_dt is declared as the first partitioned column, the country_iso2 partition is organized as subdirectories for temperature_dt. Figures 3 and 4 show the hierarchical structure of the weather_hourly_events table.

**Figure 3:** Partition organization of Apache Iceberg table weather_hourly_events by temperature_dt field.

**Figure 4:** Partition organization of Apache Iceberg table weather_hourly_events by country_iso2 field.

Since country_iso2 is at the lowest level of the hierarchy in the context of partitioning, each directory in this field contains parquet and parquet.crc files as shown in Figure 5.

**Figure 5:** Files Organization of Apache Iceberg table weather_hourly_events.

## Conclusion

The study provides a comparative evaluation of Apache Spark's performance in three programming languages: Java, Python, and Scala. For the analysis, several operations were conducted, including downloading data from CSV files, transforming the data, and loading it into Apache Iceberg tables. The results reveal several significant trends that can guide future system design.

The dependency of performance on data volume is not linear across programming languages. Python demonstrated faster execution on small and moderately sized datasets, with processing times of 6.71 seconds and 46.34 seconds, owing to its efficient high-level APIs and minimal JVM overhead. In contrast, Scala exhibited greater stability and performance in complex multi-step ETL operations involving split writes and large joins, completing the most demanding task in 374.42 seconds. This result outperformed Java (379.8 seconds) by 1.4% and Python (398.32 seconds) by 6%. These findings suggest that Python is particularly well-suited for lightweight ETL tasks involving small to medium-sized datasets, offering both high processing efficiency and ease of implementation. In contrast, Scala and Java are better suited for handling large datasets that require partitioned writes or complex joins, with Scala demonstrating the highest level of optimization for these types of workloads.

The experimental results indicate that, even under identical Spark configurations, execution time still shows variability related to language-specific execution factors such as garbage collection behavior, JVM warm-up time, and serialization efficiency. This highlights the importance of considering both Spark parameters and the execution characteristics of the programming language when optimizing large-scale ETL pipelines.

While the results were obtained in a fixed computing environment, they indicate potential areas for further research into the reliability of these observations across various resource allocations, cluster sizes, and cloud deployments. Building on the findings of this study, future research will focus on exploring similar Lakehouse storage systems, such as Delta Lake and Apache Hudi, from the perspective of developing a reliable and optimized solution for regular data loading and management. Additionally, performance benchmarking in cloud-native environments will be conducted to assess the scalability and robustness of these systems.

## Declaration on Generative AI

The authors have not employed any Generative AI tools.